\documentclass[aps,prev,twocolumn,preprintnumbers,floatfix,nofootinbib]{revtex4-1}
\pdfoutput=1
\usepackage{graphicx}
\usepackage{bm}
\usepackage{times}
\usepackage{slashed}
\usepackage{color}
\usepackage{color}
          \font\sixrm=cmr6

\newcommand{\be}{\begin{equation}}
\newcommand{\ee}{\end{equation}}
\newcommand{\bea}{\begin{eqnarray}}
\newcommand{\eea}{\end{eqnarray}}
\def\tauDM{\tau_{\hbox{\sixrm DM}}}
\def\massDM{M_{\hbox{\sixrm DM}}}
\def\rhoDM{\rho_{\hbox{\sixrm DM}}}
\def\ssection#1{\vspace{-10pt} \section{#1} \vspace{-7pt}}

\begin{document}
\begin{flushright}
MI-TH-1532
\end{flushright}

\title{New Limits on the Dark Matter Lifetime from Dwarf Spheroidal Galaxies using {\it Fermi}-LAT}

\author{Matthew G. Baring$^a$}
\email{baring@rice.edu}

\author{Tathagata Ghosh$^b$}
\email{ghoshtatha@neo.tamu.edu}

\author{Farinaldo S. Queiroz$^c$}
\email{queiroz@mpi-hd.mpg.edu}

\author{Kuver Sinha$^d$}
\email{kusinha@syr.edu}

\affiliation{(a) Department of Physics and Astronomy, Rice University, Houston, TX, 77005-1892, USA\\
(b) Department of Physics and Astronomy, Mitchell Institute for Fundamental Physics and Astronomy, Texas A\& M University,
College Station, TX 77843-4242, USA\\
(c) Max-Planck-Institut fur Kernphysik, Saupfercheckweg 1, 69117 Heidelberg, Germany\\
(d) Department of Physics, Syracuse University, Syracuse, NY 13244, USA}

\begin{abstract}
Dwarf spheroidal galaxies (dSphs) are promising targets for the indirect detection of dark matter through gamma-ray emission due to their proximity, lack of astrophysical backgrounds and high dark matter density. They are often used to place restrictive bounds on the dark matter annihilation cross section. In this paper, we analyze six years of {\it Fermi}-LAT gamma-ray data from  19 dSphs that are satellites of the Milky Way, and derive from a stacked analysis of 15 dSphs, robust 95\% confidence level lower limits on the dark matter lifetime for several decay channels and dark matter masses between $\sim 1$~GeV and $10$~TeV. Our findings are based on a bin-by-bin maximum likelihood analysis treating the J-factor as a nuisance parameter using PASS 8 event-class. Our constraints from this ensemble are among the most stringent and solid in the literature, and competitive with existing ones coming from the extragalactic gamma-ray background, galaxy clusters, AMS-02 cosmic ray data, Super-K and ICECUBE neutrino data, while rather insensitive to systematic uncertainties. In particular, among gamma-ray searches, we improve existing limits for dark matter decaying into $\bar{b}b$\, ($\mu^+\mu^-$) for DM masses below $\sim 30\, (200)$~GeV, demonstrating that dSphs are compelling targets for constraining dark matter decay lifetimes.
\end{abstract}


\maketitle

\section{Introduction} 
\vspace{-7pt}
The existence of dark matter (DM) is well established from observations of galaxies and galactic clusters,
and the cosmic microwave background, although its identity remains elusive. In the context of particle physics, DM is often interpreted as Weakly Interacting Massive Particles (WIMPs) with cross sections and masses not far from the electroweak scale. The number density of DM particles is fixed at thermal decoupling in the canonical freeze-out scenario at high redshift. The leftover DM species permeate our Universe, inducing potential signatures in deep underground experiments, colliders and astronomical telescopes/satellites. \\

DM particles do not have to be absolutely stable but simply long-lived, as happens in many well motivated theories (for an excellent review, we refer to \cite{Ibarra:2013cra}). In general the longevity of particles is attributed to the conservation of quantum numbers. For instance, in the case of standard model particles the non-observation of proton decay $p \rightarrow e^+ \pi^0$, electron decay $e \rightarrow \nu \gamma$, and neutrino decay $\nu \rightarrow \gamma\gamma$ are attributed to the conservation of  baryon number, electric charge and angular momentum, respectively. In the case of DM particles, there is no such correspondence based on fundamental symmetries. Therefore DM particles can well be stable on cosmological distance scales, with lifetimes much longer than the age of the universe  (13.8 Gyr $= 4.56 \times 10^{17}\,$sec) (see \cite{Mambrini:2015sia,Boucenna:2012rc} for a recent discussion). Such a general requirement should be quantified with no prejudice to current observations, as it has been in the context of extragalactic background radiation (EGRB) \cite{Ando:2015qda,Ackermann:2012rg,Cirelli:2012ut,Cirelli:2009dv,Abdo:2010nz}, Galaxy Clusters \cite{Huang:2011xr,Combet:2012tt,Dugger:2010ys}, anti-proton  \cite{Delahaye:2013yqa,Grefe:2015jva, Ibe:2013nka} and x-ray data \cite{Malyshev:2014xqa}, the Cosmic Microwave Background \cite{Mambrini:2015sia,Diamanti:2013bia} and optimized targets using {\it Fermi}-LAT data \cite{Massari:2015xea}. These datasets have also been used for DM annihilations.

In this paper, we set constraining limits on the DM lifetime using {\it Fermi}-LAT gamma-ray data from the observation of dSphs. dSphs that are proximate to the Milky Way are special targets for indirect detection of DM signals for several reasons: (i) their gravitational dynamics indicate that they are DM-dominated objects; (ii) they are generally located at moderate or high Galactic latitudes and therefore are subject to low diffuse gamma-ray foregrounds; (iii) their lack of unambiguously discernible astrophysical gamma-ray emission; (iv) they possess relatively small uncertainties on the DM profile. Thus, it is fruitful to derive bounds on DM properties using dSphs observations. 

A first offering of dSphs constraints on DM lifetimes was made in \cite{Dugger:2010ys} using around one year of {\it Fermi}-LAT observations \footnote{See \cite{Malyshev:2014xqa} dSphs studies for the keV line.}. Yet greater emphasis in the literature has been on constraining DM annihilation cross sections. In \cite{PalomaresRuiz:2010pn}, the authors focused on how to distinguish a signal coming from DM annihilation and/or decay using dSphs observations from gamma-ray experiments, whereas in \cite{Colafrancesco:2006he} a multi-wavelength approach was performed for annihilating DM, and in \cite{Gonzalez-Morales:2014eaa} the impact of hosting intermediate massive black holes was investigated. Various aspects of DM annihilation in these contexts were explored in \cite{Dutta:2015ysa}.
The {\it Fermi}-LAT collaboration has invested extensive effort in increasing the sensitivity to potential DM signals \cite{Ackermann:2013yva,Ackermann:2015abc}, including updates to their point source catalog, and upgrades to the event reconstruction and foreground/background subtraction afforded by the new 
PASS 8 analysis tool.  These have resulted in stringent bounds on the annihilation cross section \cite{GeringerSameth:2011iw}. \\

For dark matter decay, here we extend and complement previous works by including six years of LAT data and also employing the new PASS8 event class. Moreover, for the first time in the literature for decay studies, we stack a larger pool of 15 dSphs using a bin-by-bin maximum likelihood method, treating the astrophyiscal $J$-factor of the dSphs as nuisance parameters.  This protocol renders our conclusions robust, and less sensitive to systematics and statistical uncertainties. The baseline conclusion is that herein we raise the decay liftetime lower bounds of \cite{Dugger:2010ys} by factors of around 3-10. \\

For our focus on dark matter decays, the gamma-ray flux (see Eq.~[\ref{eq:flux}]) from any DM congregation is linearly proportional to the J-factor $J_{\rm d}$ (see Eq.~[\ref{eq:Jfac_def}]) for the volume-integrated DM content of a galaxy.  The J-factors for dSphs are fairly accurately estimated: recent measurements of the stellar velocity dispersion and half-light radius have led to better determinations of these J-factors \cite{Walker:2009zp,Wolf:2009tu,Martinez:2013els}, and such improvements are exploited here to define more accurate bounds on DM properties.\\

\begin{figure*}[!t]
\mbox{\includegraphics[width=\columnwidth]{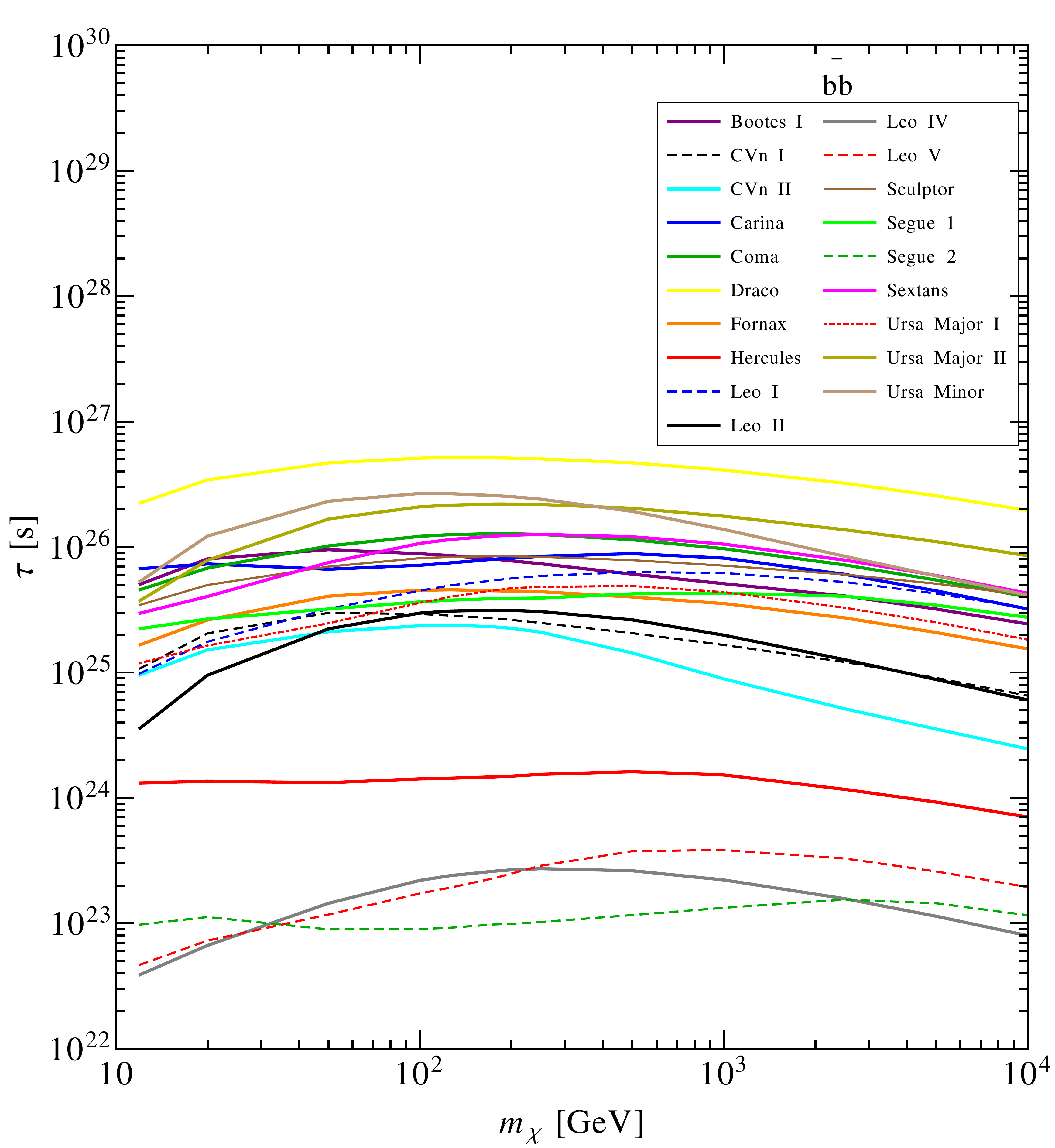}\quad\includegraphics[width=\columnwidth]{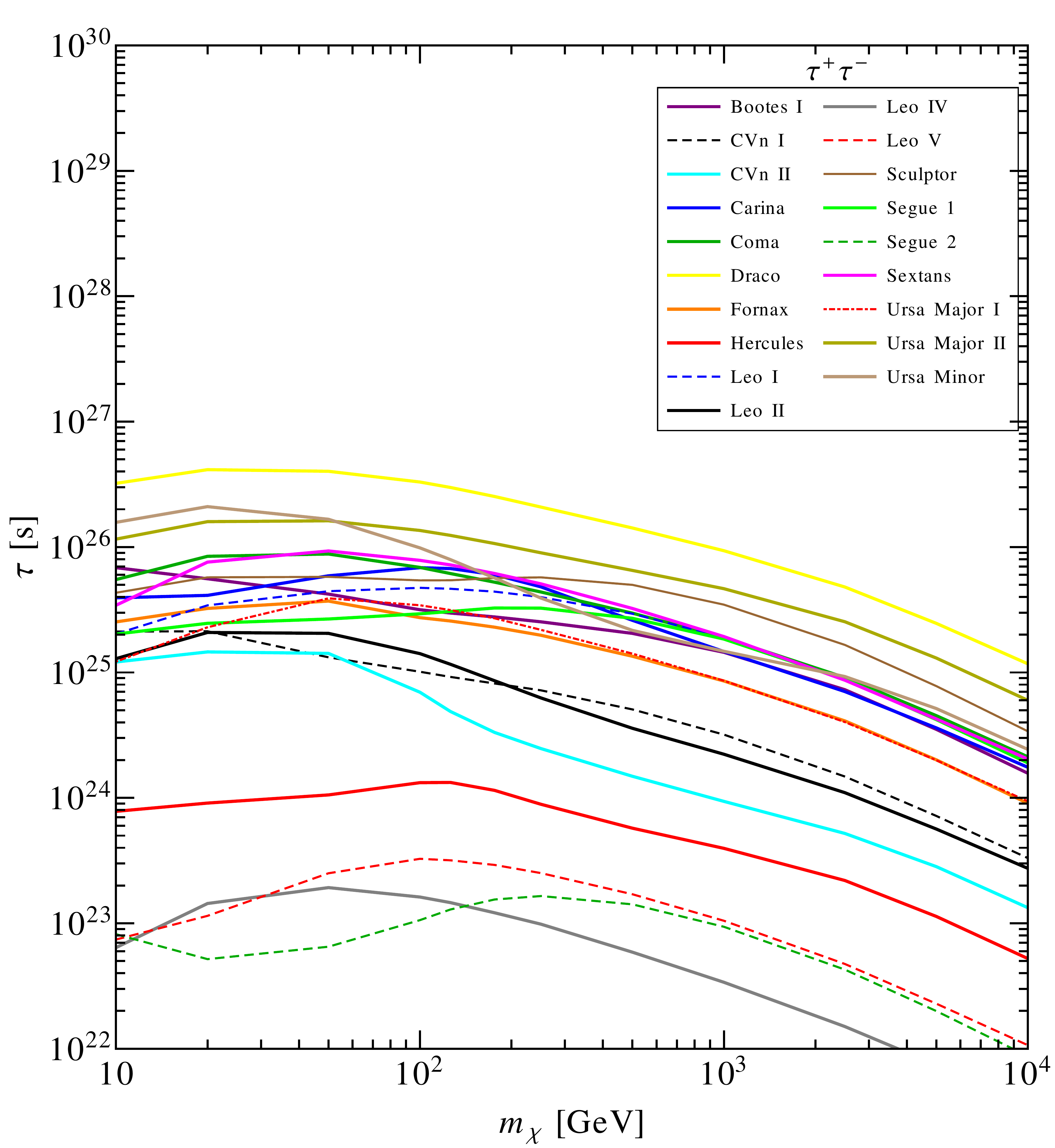}}
\vspace{-10pt}
\caption{$95\%$ C.L individual lower limits on the dark matter lifetime from the most constraining dSphs for the bb (left panel) and $\tau\tau$ (right panel) final state. We explicitly derived individual limits from all 19 dSphs and decided to plot the ones which yield the most restrictive bounds for clarity. It is clear that Draco, Ursa Minor and Ursa Major II provide the best limits and are in the ballpark of $10^{26-27}$~sec.}
\label{Limits1}
\end{figure*}

We combine these updated $J_{\rm d}$ values with extensive datasets from six years of 
observations of dSphs using {\it Fermi}-LAT. Several dSphs observed by {\it Fermi}-LAT do not have 
their J-factor estimated and are removed from our analysis. For a similar reason we are not including the new dSphs observed by Dark Energy
Survey and Panoramic Survey Telescope and Rapid Response System \cite{Bechtol:2015cbp,Koposov:2015cua,Laevens:2015una,DESCollaboration}.\\ 

\begin{figure}[!t]
\mbox{\includegraphics[width=0.9\columnwidth]{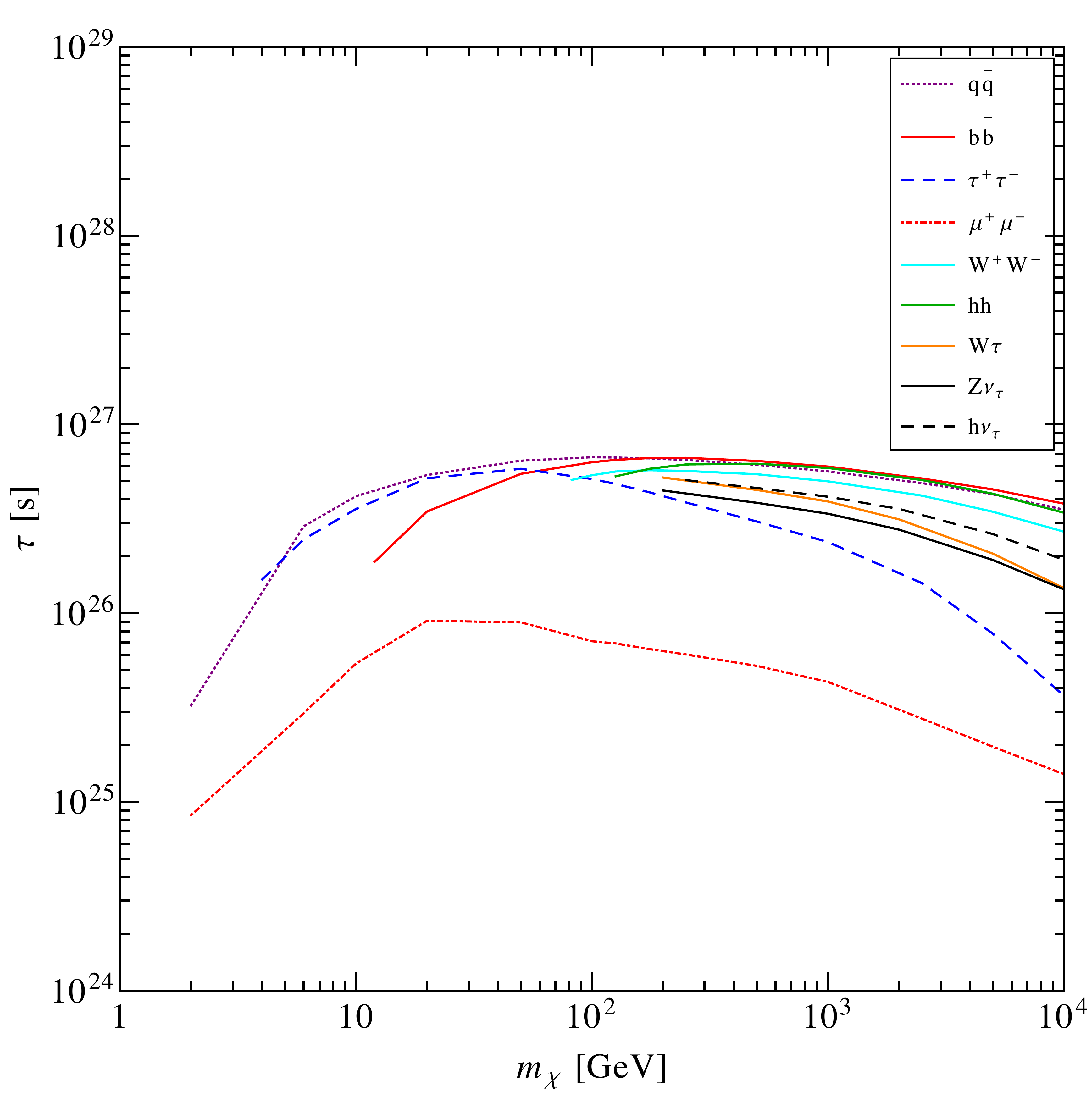}}
\vspace{-10pt}
\caption{$95\%$~C.L stacked analysis bound on DM lifetime for several decay channels, encompassing both fermionic and bosonic DM. Refer to text for the list of dSphs used in the stacked analysis. The decay into $qq$ takes into account all light quarks. For heavy DM, decays into $\bar{b}b$,$hh$ and $qq$ provide the strongest limits, whereas for relatively light dark matter $\bar{b}b$,$qq$ and $\tau\tau$ are dominant. As we shall see further in Fig. 3 these are the most stringent limits in the literature for DM from gamma-ray searches masses below  $\sim 30$~GeV ($200$~GeV) for decays into bb ($\mu^+\mu^-$).}
\label{Limits1}
\end{figure}

\ssection{Data Analysis}
We gather six years of {\it Fermi}-LAT gamma-ray data belonging to the $P8R2SOURCEV6$ instrument response function, dating since August 4, 2008, for the 19 dSphs shown in the main portion of Table I. The energy bins range from $500$~MeV to $500$~GeV. We use the Pass-8 event class which contains an improved point-spread function (PSF) and increased telescope effective area compared to previous {\it Fermi}-LAT analysis protocols. We also employ data from the new point source {\it Fermi-LAT} catalog, 3FGL.  The lower energy bound is chosen to avoid systematics due to the leakage of photons coming from the Earth limb due to poor/broad PSF  at energies lower than 500 MeV \footnote{For a list of the {\it Fermi}-LAT tools used see \url{http://fermi.gsfc.nasa.gov/ssc/data/analysis/scitools/overview.html}}.\\

As aforementioned, we show the 19 dSphs of interest plus Reticulum II  with their respective positions, distances and J-factors in Table I. Within $2\sigma$, the DM profile of all dSphs are well described by a NFW profile (see Table IV of \cite{Geringer-Sameth:2014yza}). We singled out these 19 dSphs because several dSphs, namely Bootes II, Bootes III, Canis Major, Pisces II, and Sagittarius, have J-factors that are either poorly constrained, or are not determined at all. They are thus excluded from our study. Moreover, in our stacked analysis, Canes Venatici I and Leo I were left out because their regions of interest (ROIs) in the sky overlap with Canes Venatici II and Segue 1 that have larger J-factors. Furthermore the ROI of Ursa Major I overlaps with that of Wilman 1, as pointed out in \cite{Ackermann:2013yva}. Nevertheless, Wilman 1 is omitted here since \cite{Geringer-Sameth:2014yza} did not report its J-factor. Those choices concur with those from Fermi-LAT collaboration in \cite{Ackermann:2015abc}. Hence, to avoid statistical interference and follow the procedure in \cite{Ackermann:2015abc}, we use 15 dSphs, namely Bootes I, Canes Venatici II, Carina, Coma Berenices, Draco, Fornax, Hercules, Leo II, Leo IV, Sculptor, Segue 1, Sextans, Ursa Major I, Ursa Major II, and Ursa Minor in the stacked analysis.\\

As usual, we reject events with rocking angle larger than $100^{\circ}$ to minimize contamination from the bright limb of the Earth as well as events during periods when the rocking angle of the LAT instrument was larger than $52^{\circ}$ using the {\it gtmktime} tool of {\it Fermi}-LAT software.
After defining the ROI  as in \cite{Ackermann:2015abc} with $0.1^{\circ}$ pixels and 24 energy bins logarithmically separated using {\it gtltcube} and {\it gtexpcube2} tools, we model the diffuse and isotropic background emission using the galactic and extragalactic models provided in \cite{Fermiback}. \\

We perform a bin-by-bin likelihood analysis of the gamma-ray emission within $5^{\circ}$ of each dSph galaxy's center, which set the normalizations of the diffuse sources and the normalizations of point-like background sources within $5^{\circ}$ of each dSph center as in \cite{Ackermann:2015abc}. For each dSph, the spatial DM distribution is modeled by a NFW dark matter profile with a J-factor ($J_{\rm d}$) defined as
\begin{equation}
   J_{\rm d} \; =\; \int_{\Delta\Omega} d\Omega \int \rhoDM (s)\, ds
   \quad ,\quad s\; =\; s(\theta )\quad ,
 \label{eq:Jfac_def}
\end{equation}
where the DM density $\rhoDM$ is integrated along lines of sight elements $ds$ for different directions within the ROI solid angle $\Delta\Omega$.  Values for $J_{\rm d}$ for our dSphs sample are listed in Table I, taken from \cite{Geringer-Sameth:2014yza}; these are proportional to the expected intensity of $\gamma$-ray emission from DM decay in a given ROI assuming a spherically symmetric NFW DM density distribution,
\begin{equation}
   \rho(r) \; =\; \frac{\rho_s}{r/r_s (1+ r/r_s)}\quad .
 \label{eq:DM_profile}
\end{equation}where $\rho_s$ and $r_s$ are the characteristic density and scale
radius, which are determined dynamically from the maximum circular velocity $v_c$ and the enclosed mass contained up to the radius of maximum $v_c$ \cite{Dwarfparameters}.\\

 We emphasize that within $0.5^{\circ}$ the integrated J-factor is rather insensitive to the choice of the DM density profile for slopes not steeper than 1.2 \cite{Strigari:2007at}. The integrated J-factors of our selected dSphs were obtained over a cone of radius $\theta=0.5^{\circ}$, i.e., accounting for $50\%$ of the total DM emission, which is a conservative approach. If we had instead used the larger value $\theta_{max}$ from \cite{Geringer-Sameth:2014yza}, our limits would be raised by a factor of two or so. We then compute the likelihood of an individual target i,\\
\begin{equation}
   \tilde{\mathcal{L}_i}(\mu,\theta_i = 
   \lbrace\alpha_i,J_i\rbrace |D_i)=\mathcal{L}_i(\mu,\theta_i|D_i)\mathcal{L}_J (J_i|J_{obs,i},\sigma_i)
\end{equation}
where $\mu$ are the parameters of the DM model, i.e. the product of the dark matter lifetime and mass as we shall see further, $\theta_i$ is the set of nuisance parameters that includes both nuisance
parameters from the LAT analysis ($\alpha_i$) and the dSph J-factor $J_i$, and $D_i$ is the gamma-ray data. Notice that we incorporated a likelihood J-factor term as an attempt to account for statistical uncertainties on J-factors of each dSphs which is defined as,\\

\begin{eqnarray}
\mathcal{L}_J (J_i| J_{obs,i}, \sigma_i) &= & \frac{1}{ln(10)J_{obs,i}\sqrt{2\pi} \sigma_i}\nonumber\\
\label{eq:}\\
& \times  & \exp \left\{ -\frac{(log_{10}(J_i)-log_{10}(j_{obs,i}))^2}{2\sigma_i^2} \right\}\nonumber
\end{eqnarray}
where $J_i$ is the true value of the J-factor of a dSphs $i$, and $J_{obs,i}$ is
the measured J-factor with error $\sigma_i$. We later join the likelihood functions,
\begin{eqnarray}
\mathcal{L}_i (\mu, \theta_{i}| D_i) = \prod_j \mathcal{L}_i (\mu, \theta_i | D_{i,j})
\label{Maxlike}
\end{eqnarray} 

Notice that this procedure, which matches the one adopted in \cite{Ackermann:2015abc}, is independent of the DM energy spectrum  in each energy bin, since it corresponds to an upper limit on the energy flux. We now evaluate the test statistic (TS) defined as $TS= -2 ln ( \mathcal{L} (\mu_0,\widehat{\theta}|D)/ \mathcal{L} (\widehat{\mu},\widehat{\theta}|D))$, and require a change in the profile log-likelihood of $=2.71/2$ from its maximum corresponding to 95\% C.L. upper limit on the energy flux as described in \cite{Rolke:2004mj}. In the next section we discuss the expected gamma-ray signal from DM decay and our results based on the aforementioned procedure.\\

\begin{table}
Nearby Dwarf Spheroidal Galaxies
\begin{tabular}{c|c|c|c|c}
\hline
Name & l & b & distance & $\log_{10}J_{\rm d} (\theta_{0.5})$ \, {\rm GeV cm$^{-2}$}\\
\hline
Bootes I & 358.1 & 69.6 & 66 kpc & $17.90^{+0.23}_{-0.26}$\\
Canes Venatici I & 74.3 & 79.8 & 218 & $17.57^{+0.36}_{-0.72}$\\
Canes Venatici II & 113.6 & 82.7 & 160 & $16.97^{+0.24}_{-0.23}$\\
Carina & 260.1 & -22.2 & 105 & $17.90^{+0.17}_{-0.16}$\\
Coma Berenices & 241.9 & 83.6 & 44 & $17.96^{+0.20}_{-0.25}$\\
Draco & 86.4 & 34.7 & 76 & $18.53^{+0.10}_{-0.12}$\\
Fornax & 237.1 & -65.7 & 147 & $17.86^{+0.04}_{-0.05}$\\
Hercules & 28.7 & 36.9 & 132 & $16.66^{+0.42}_{-0.40}$\\
Leo I & 226.0 & 49.1 & 254 & $17.91^{+0.15}_{-0.20}$\\
Leo II & 220.2 & 67.2 & 233 & $17.24^{+0.35}_{-0.48}$\\
Leo IV & 265.4 & 56.5 & 154 & $16.12^{+0.71}_{-1.14}$\\
Leo V  & 261.9 & 58.5 & 178 &  $15.86^{+0.46}_{-0.47}$ \\
Sculptor & 287.5 & -83.2 & 86 & $18.19^{+0.07}_{-0.06}$ \\
Segue 1 & 220.5 & 50.4 & 23 & $17.99^{+0.20}_{-0.31}$\\
Segue 2 & 149.4 & -38.1 & 35 & $15.89^{+0.56}_{-0.37}$\\
Sextans & 243.5 & 42.3 & 86 & $17.89^{+0.13}_{-0.23}$\\
Ursa Major I & 159.4 & 54.4 & 97 & $17.61^{+0.20}_{-0.38}$\\
Ursa Major II & 152.5 & 37.4 & 32 & $18.38^{+0.25}_{-0.27}$\\
Ursa Minor & 105.0 & 44.8 & 76 & $18.03^{+0.16}_{-0.13}$ \\
\hline
Reticulum II & 266.3 & -49.7 & 32 & $18.8^{+0.7}_{-0.7}$\\
\hline
\end{tabular}
\caption{Galactic longitude ($l$), latitude ($b$), distance (in kpc), and DM decay J-factor 
for 20 dwarf galaxies that are satellites of the Milky Way. The $J_d$ factors are integrated over a cone of radius $\theta_{0.5}$, where $\theta_{0.5}$ is the "half-life radius" i.e. the angle containing $50\%$ of the total dark matter emission. For Reticulum II,  we adopted the J-factor value reported in \cite{Bonnivard:2015tta}. For other dwarf galaxies, we adopted the values reported in \cite{Geringer-Sameth:2014yza}; see text for details.}  
\end{table}


\ssection{Lower Bounds to Dark Matter Lifetimes}
The differential flux of photons from a given angular direction $\Delta\Omega$ within an ROI 
produced by the decay of a DM particle into a single final state is expressed as
\begin{equation}
\label{eq:flux}
\Phi_{\gamma} (\Delta \Omega) \; =\; \frac{1}{4 \pi \massDM \tauDM } \int_{E_{min}}^{E_{max}}
\frac{dN_{\gamma}}{dE_{\gamma}} dE_{\gamma} \,\cdot\, J_{\rm d}\;\; ,
\end{equation}
where $\massDM, \tauDM$ and $dN_{\gamma}/dE_{\gamma}$ are the DM mass, lifetime and differential $\gamma$-ray yield per decay, respectively. In a given particle physics model, in order to find the total gamma-ray flux coming from the decay of a DM particle, $dN_{\gamma}/dE_{\gamma}$ has to be summed over all possible final states. In this work, however, we focus on one final state channel at a time, and compute the energy spectrum using PPPC4DM \cite{Cirelli:2010xx} for the $q\bar{q}$, $b\bar{b}$,$\tau^+\tau^-$,$\mu^+\mu^-$,$W^+W^-$ and $hh$, and instead the Pythia code \cite{Sjostrand:2007gs} for the $Z\nu_{\tau}$,$h\nu_{\tau}$ and $W\nu_{\tau}$ channels. With the energy spectrum determined, we can compute the profile likelihood function for the lifetime $\tauDM$ vs $\massDM$ by maximizing the global likelihood function in Eq.~(\ref{Maxlike}) with respect to the nuisance parameters and derive our bounds.\\

In the left panel of Fig.1, we exhibit the constraints on the DM lifetime for decays into $\bar{b}b$ for the 19 dSphs in our study. Draco, Ursa Minor and Ursa Major II give rise to the strongest bounds on the DM lifetime due to their proximity and their large J-factors. Draco excludes a DM lifetime smaller than $\sim 2\times 10^{26}$~sec (i.e., $> 10^8$ Hubble times) at $95\%$ C.L. for DM masses below 10 TeV. The characteristic mass dependence of the limit curves for each individual dSph can be explained by comparing the shape of the upper flux limit curve and the energy spectrum of the final decay state, in this case $b\bar{b}$. The strongest bounds from the upper flux occur at energies of few GeV or so, which roughly coincides with the peak in the $b\bar{b}$ energy spectrum for dark matter masses 10GeV-10TeV.\\

In the right panel, we exhibit the limits on decays into $\tau\tau$ pairs for the same dSph set. For DM masses below $100$~GeV we found a lower limit of $\tauDM \, \sim \, 3\times 10^{26}$~sec at $95\%$ C.L. For such masses, decays into $\tau^+\tau^-$ produce more photons than those from $b\bar{b}$, and this 
leads to the slight skewing of the limit curves.\\

What we meant is that the dark matter distribution of each of the dSphs does not have 
to be precisely the same. Plus, the upper limit on the flux of each individual dSph differs, 
which results in different limits on the dark matter lifetime. In other words, for a given final state, 
the shape of the limit curve differs from galaxy to galaxy, as can be seen in Fig.1. 
Therefore, stacking a large sample of dSphs makes our combined limit less sensitive to a 
peculiar dwarf galaxy, i.e. more broadly representative.

The use of an individual dSph to place constraints on DM properties might bias bounds, since the dark matter distribution of each of the dSphs does not have to be precisely the same. Plus, the upper limit on the flux of each individual dSph differs, which results in different limits on the dark matter lifetime.  In other words, for a given final state, 
the shape of the limit curve differs from galaxy to galaxy, as can be seen in Fig.1. 
Therefore, stacking a large sample of dSphs makes our combined limit less sensitive to a 
peculiar dwarf galaxy, yet more broadly representative.  For these reasons we performed a maximum likelihood analysis of a stack of 15 dSphs treating the J-factors as nuisance parameters, as described in the previous section, and obtained stringent constraints on the DM lifetime shown in Fig.2 for the decay modes: $ hh, h\nu, WW, W\tau, Z\nu,bb, \mu\mu, \tau\tau$ and $qq$, where $qq$ includes decays into all light quarks. These decay channels encompass both fermionic and bosonic DM, making our bounds applicable to a plethora of DM models. \\

As expected, after properly stacking the bin-by-bin likelihood functions of each dSphs, our combined limit gets stronger and less sensitive to systematic and statistical uncertainties. We conclusively exclude decay lifetimes up to $7 \times 10^{26}$~sec into $b\bar{b}$ and $4 \times 10^{26}$~sec into $\tau^+\tau^-$.  Most of the final states have a kinematic cut-off prohibiting existence of limits for certain masses. For sufficiently small DM masses most of the photons appear outside the energy window of interest (i.e. below 500 MeV), thereby defining the sharp drop for channels such as $\mu^+\mu^-,\tau^+\tau^-$ and $q\bar{q}$.\\

For instance, for fermionic DM, such as the gravitino in supersymmetry \cite{Bolz:2000fu,Pradler:2006qh}, decays into $h\nu_{\tau}$, $Z\nu_{\tau}$, $W\tau$ are limited to lifetimes larger than $1-3\times 10^{26}$~sec. In the context of supersymmetric grand unification, six dimension mass operators may lead to DM decay with a lifetime $\tau_{\rm susy} \simeq 6.3\,\times 10^{25}\, {\rm sec}\, (\Lambda_{\rm GUT}/10^{16} {\rm GeV})^4\,({\rm TeV}/\massDM)^5$ \cite{Dugger:2010ys}. For decay into bb, we bound the scale of unification ($\Lambda_{\rm GUT}$) to be larger than $\sim 10^{15}\; {\rm GeV} \; (\sim 2.8\,\times 10^{17}\, {\rm GeV})$ for a $100$~GeV ($10$~TeV) DM particle. \\

\begin{figure*}[!t]
\mbox{\includegraphics[width=0.92\columnwidth]{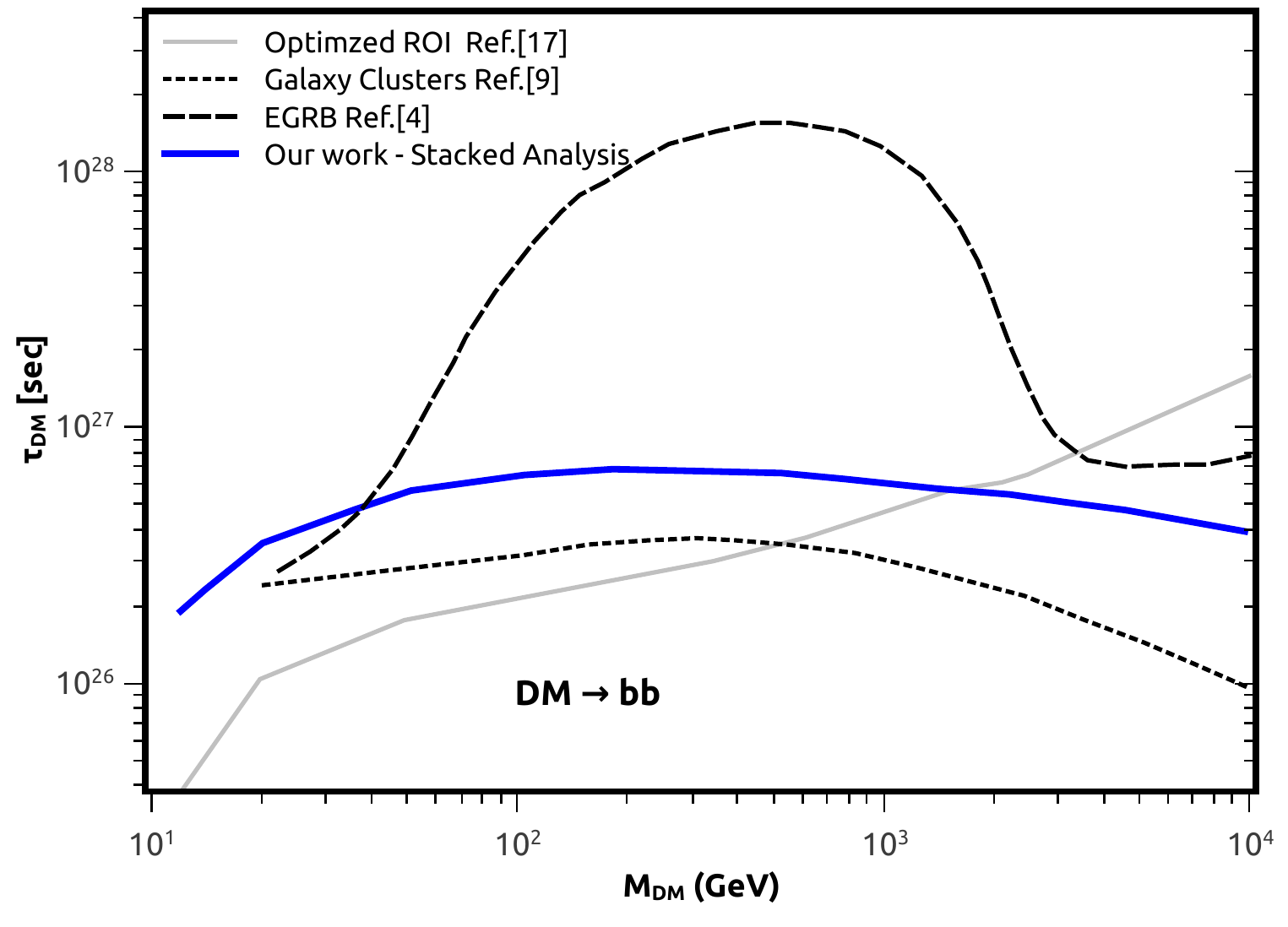}\quad\includegraphics[width=0.90\columnwidth]{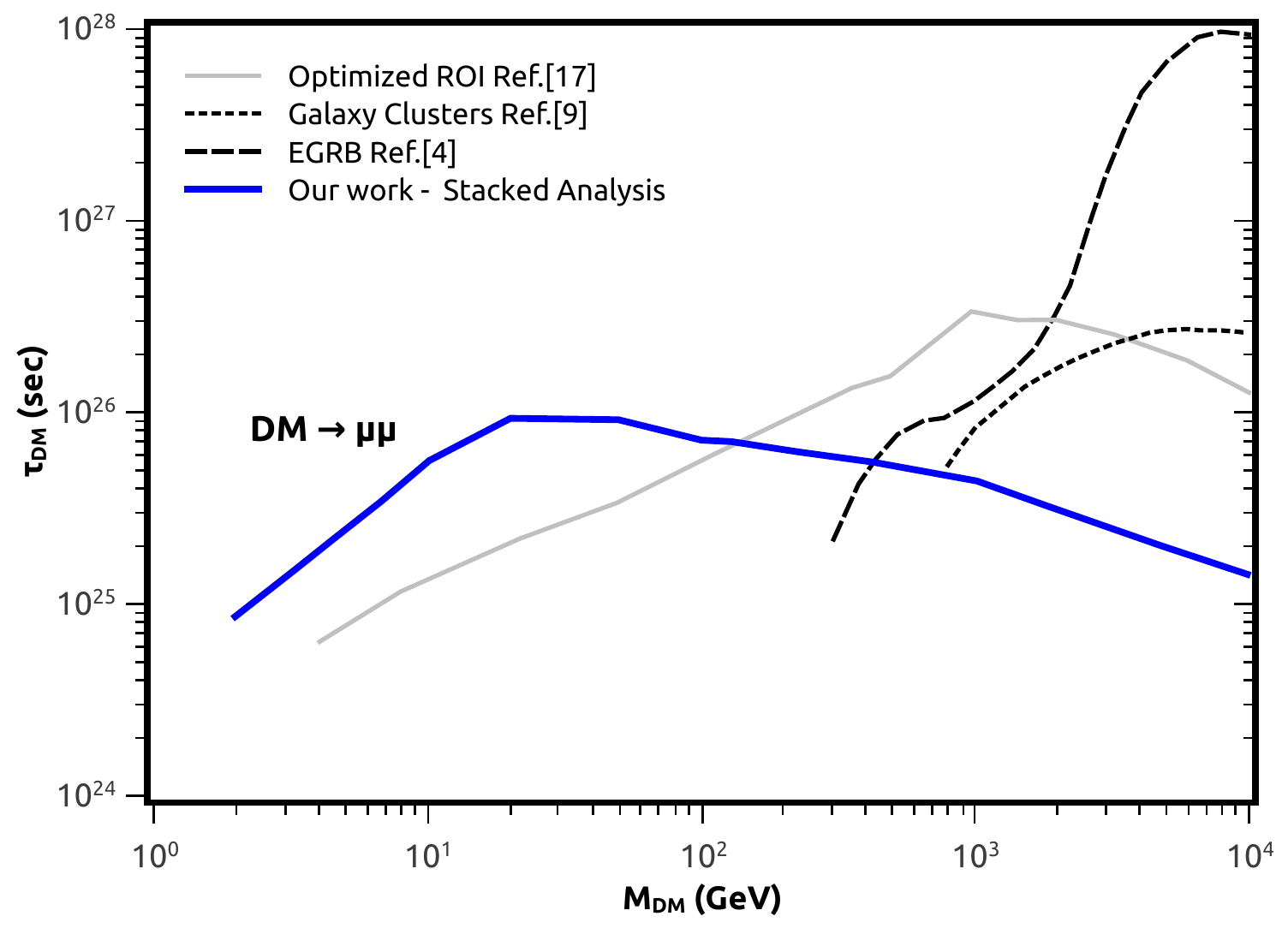}}
\vspace{-10pt}
\caption{In both plots we compare our results with existing gamma-ray observations, namely bounds extragalactic gamma-ray background (EGRB) from \cite{Ando:2015qda} (dashed) and galaxy clusters from \cite{Huang:2011xr} (dotted) and using optimized ROI strategy \cite{Massari:2015xea} (solid gray) employing {\it Fermi}-LAT data. See text for detail. {\it Left}: limits for dark matter decay into $\bar{b}b$; {\it Right} limits for dark matter decay into $\mu^+\mu^-$. We conclude that our limits are the strongest for dark matter masses below $\sim 30$~GeV and $\sim 200$~GeV for the bb and $\mu^+\mu^-$ decay channels respectively.}
\label{Limits1}
\end{figure*}

These findings demonstrate that stacked studies of dSphs provide robust and restrictive lower limits for the DM lifetime. To provide context for our study, it is insightful to compare our dSph bounds with constraints from various other gamma-ray searches for decaying DM. To facilitate this, in the left panel of Fig.3 we gather limits from different gamma-ray search strategies. There we plotted the limits coming from extragalactic gamma-ray background (EGRB) derived in \cite{Ando:2015qda} (Fig.3 foreground model A) with a dashed line, and galaxy clusters  \cite{Huang:2011xr} (Fig.4 for $b\bar{b}$ and Fig.5 for $\mu^+\mu^-$) with a dotted black line, optimized ROI searches \cite{Massari:2015xea} with solid gray line, along with our limits from a stacked analysis (blue curve). For the $b\bar{b}$ channel, our bounds improve upon previous results for dark matter masses below 30 GeV or so, whereas for decays into $\mu^+\mu^-$, our constraints are the most restrictive for masses below $\sim 200$~ GeV. 

One should keep in mind that \cite{Huang:2011xr} used older data and Fermi-LAT software, therefore an improvement on their limit is expected when updating the data/analysis specially for the bb final state, though it is beyond the scope of our manuscript to compute it. Here, we simply quote their results.

We stress that antiproton (positron fraction) data may provide stronger limits \cite{Mambrini:2015sia,Giesen:2015ufa,Audren:2014bca}  on DM decaying into $\bar{b}b$ ($\mu^+\mu^-$), but since these are subject to rather large uncertainties we left them out, and focused our comparison among gamma-ray searches. Moreover, we neglected existing limits from PLANCK data \cite{Mambrini:2015sia,Diamanti:2013bia}, Super-K and ICECUBE \cite{Aisati:2015vma} on $\mu^+\mu^-$ since they are much weaker.\\

\begin{figure}[!h]
\mbox{\includegraphics[width=0.9\columnwidth]{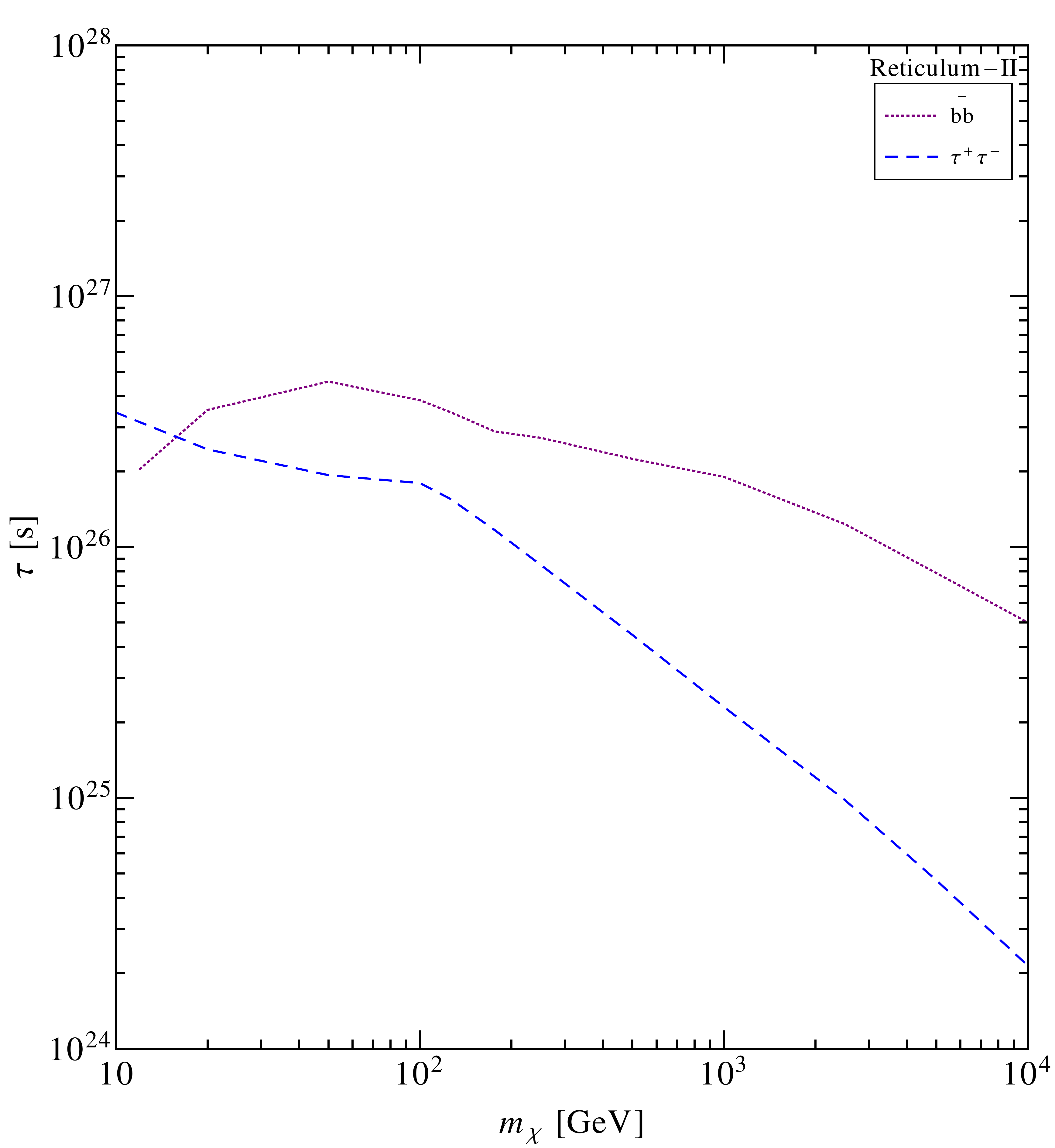}}
\vspace{-10pt}
\caption{$95\%$~C.L bound on DM lifetime for $b\bar{b}$ and $\tau^+\tau^-$ channels using Reticulum-II data only. Notice the anomalous behavior for $b\bar{b}$, which provides a limit still weaker than our combined one. As for the other dSphs the aspect of the limit curves are result of a combination of both energy spectrum of the final state and the upper flux limit reported in \cite{Drlica-Wagner:2015xua}.}
\label{Limitsreti}
\end{figure}

We point out that recently a $\gamma$-ray excess has been claimed for a newly discovered dwarf galaxy Reticulum II \cite{Geringer-Sameth:2015lua}. The {\it Fermi}-LAT collaboration has independently performed a similar analysis that indicates that the excess above $\sim 100$ MeV  is merely a statistical fluctuation of the background, since no surplus of photons is observed in the remaining dwarf galaxies \cite{Drlica-Wagner:2015xua,Hooper:2015ula}. The origin of this $\gamma$-ray emission is unclear, especially because the two groups used different datasets, and their conclusion concerning the chance of a background fluctuation mimicking the potential dark matter signal differs. For these reasons we have omitted Reticulum II from the stacked analysis, but as a contextual note we obtained the limits on the dark matter lifetime, exhibited in Fig.\ref{Limitsreti}, stemming from Reticulum II using the upper flux reported in \cite{Drlica-Wagner:2015xua} with the J-factor presented in the Table I.

It is clear that there is anomalous behavior for $b\bar{b}$, which provides a very good fit to the excess seen in \cite{Geringer-Sameth:2015lua} for DM masses around dozens of GeV. The limit arising from Reticulum-II is still weaker than our combined one. Similar to the other dSphs, the shapes of the limit curves result from a combination of the shape of the energy spectrum of the final state, and the upper flux reported in \cite{Drlica-Wagner:2015xua}.\\

\ssection{Conclusions} 
In this paper, we have used 500MeV-500GeV gamma-ray data from the {\it Fermi}-LAT observation of Milky Way satellite dSphs to place stringent and robust lower bounds to the DM lifetime. We derived individual and stacked limits for several channels for the first time in the literature. We further compared our results with others from different search strategies, conclude that among gamma-ray searches dSphs are the leading ones for dark matter  masses below $30$~GeV and $200$~GeV for the $\bar{b}b$ and $\mu^+\mu^-$ final states, respectively.  Our findings show that gamma-ray searches from the observation of dSphs using {\it Fermi}-LAT data are compelling targets for probing dark matter decay physics. \\

{\it Acknowledgement}\\

We thank the Referees for their comments which led to the inclusion of a maximum likelihood analysis and PASS8 event-class. The authors thank Louie Strigari, Stefano Profumo, Marco Cirelli, Chistoph Weniger, Alessandro Ibarra, Celine Boehm, Manfred Lindner, Brandon Anderson, Carlos Yaguna and Joseph Silk for correspondence/discussion. F.S.Q. thanks HAP Dark Matter meeting for providing an enlightening environment during final stages of the paper. T.G. is supported by DOE Grant DE-FG02-13ER42020. K.S. is supported by NASA Astrophysics Theory Grant NNH12ZDA001N.


\begin{thebibliography}{99}

\bibitem{Ibarra:2013cra} 
  A.~Ibarra, D.~Tran and C.~Weniger,
  Int.\ J.\ Mod.\ Phys.\ A {\bf 28}, 1330040 (2013)
  [arXiv:1307.6434 [hep-ph]].

\bibitem{Mambrini:2015sia} 
  Y.~Mambrini, S.~Profumo and F.~S.~Queiroz,
  arXiv:1508.06635 [hep-ph].


\bibitem{Boucenna:2012rc} 
  M.~S.~Boucenna, R.~A.~Lineros and J.~W.~F.~Valle,
  Front.\ Phys.\  {\bf 1}, 34 (2013)
  [arXiv:1204.2576 [hep-ph]];
P.~Fileviez Perez,
  arXiv:1501.01886 [hep-ph];
W.~Rodejohann and C.~E.~Yaguna,
  arXiv:1509.04036 [hep-ph].  
\bibitem{Ando:2015qda} 
  S.~Ando and K.~Ishiwata,
  JCAP {\bf 1505}, no. 05, 024 (2015)
  [arXiv:1502.02007 [astro-ph.CO]].

  
  \bibitem{Ackermann:2012rg} 
  M.~Ackermann {\it et al.} [{\it Fermi}-LAT Collaboration],
  Astrophys.\ J.\  {\bf 761}, 91 (2012)
  [arXiv:1205.6474 [astro-ph.CO]].


\bibitem{Cirelli:2012ut} 
  M.~Cirelli, E.~Moulin, P.~Panci, P.~D.~Serpico and A.~Viana,
  Phys.\ Rev.\ D {\bf 86}, 083506 (2012)
  [arXiv:1205.5283 [astro-ph.CO]].
  
\bibitem{Cirelli:2009dv} 
  M.~Cirelli, P.~Panci and P.~D.~Serpico,
  Nucl.\ Phys.\ B {\bf 840}, 284 (2010)
  [arXiv:0912.0663 [astro-ph.CO]].
 

\bibitem{Abdo:2010nz} 
  A.~A.~Abdo {\it et al.} [{\it Fermi}-LAT Collaboration],
  Phys.\ Rev.\ Lett.\  {\bf 104}, 101101 (2010)
  [arXiv:1002.3603 [astro-ph.HE]].
  
\bibitem{Huang:2011xr} 
  X.~Huang, G.~Vertongen and C.~Weniger,
  JCAP {\bf 1201}, 042 (2012)
  [arXiv:1110.1529 [hep-ph]].

\bibitem{Combet:2012tt} 
  C.~Combet, D.~Maurin, E.~Nezri, E.~Pointecouteau, J.~A.~Hinton and R.~White,
  Phys.\ Rev.\ D {\bf 85}, 063517 (2012)
  [arXiv:1203.1164 [astro-ph.CO]].

  \bibitem{Dugger:2010ys} 
  L.~Dugger, T.~E.~Jeltema and S.~Profumo,
  JCAP {\bf 1012}, 015 (2010)
  [arXiv:1009.5988 [astro-ph.HE]].

\bibitem{Delahaye:2013yqa} 
  T.~Delahaye and M.~Grefe,
  JCAP {\bf 1312}, 045 (2013)
  [arXiv:1305.7183 [hep-ph]].

\bibitem{Grefe:2015jva} 
  T.~Delahaye and M.~Grefe,
  JCAP {\bf 1507}, no. 07, 012 (2015)
  [arXiv:1503.01101 [hep-ph]].

\bibitem{Ibe:2013nka} 
  M.~Ibe, S.~Iwamoto, S.~Matsumoto, T.~Moroi and N.~Yokozaki,
  JHEP {\bf 1308}, 029 (2013)
  [arXiv:1304.1483 [hep-ph]].

\bibitem{Malyshev:2014xqa} 
  D.~Malyshev, A.~Neronov and D.~Eckert,
  Phys.\ Rev.\ D {\bf 90}, 103506 (2014)
  doi:10.1103/PhysRevD.90.103506
  [arXiv:1408.3531 [astro-ph.HE]].
\bibitem{Diamanti:2013bia} 
  R.~Diamanti, L.~Lopez-Honorez, O.~Mena, S.~Palomares-Ruiz and A.~C.~Vincent,
  JCAP {\bf 1402}, 017 (2014)
  [arXiv:1308.2578 [astro-ph.CO]].

\bibitem{Massari:2015xea} 
  A.~Massari, E.~Izaguirre, R.~Essig, A.~Albert, E.~Bloom and G.~A.~Gómez-Vargas,
  Phys.\ Rev.\ D {\bf 91}, no. 8, 083539 (2015)
  [arXiv:1503.07169 [hep-ph]].

\bibitem{PalomaresRuiz:2010pn} 
  S.~Palomares-Ruiz and J.~M.~Siegal-Gaskins,
  JCAP {\bf 1007}, 023 (2010)
  [arXiv:1003.1142 [astro-ph.CO]].  

\bibitem{Colafrancesco:2006he} 
  S.~Colafrancesco, S.~Profumo and P.~Ullio,
  Phys.\ Rev.\ D {\bf 75}, 023513 (2007)
  [astro-ph/0607073].  

\bibitem{Dutta:2015ysa} 
  I.~Cholis and P.~Salucci,
  Phys.\ Rev.\ D {\bf 86}, 023528 (2012)
  doi:10.1103/PhysRevD.86.023528
  [arXiv:1203.2954 [astro-ph.HE]];

  B.~Dutta, Y.~Gao, T.~Ghosh and L.~E.~Strigari,
  arXiv:1508.05989 [hep-ph];
   M.~A.~Fedderke, E.~W.~Kolb, T.~Lin and L.~T.~Wang,
  JCAP {\bf 1401}, no. 01, 001 (2014)
  [arXiv:1310.6047 [hep-ph]];
  N.~Fernandez, J.~Kumar, I.~Seong and P.~Stengel,
  Phys.\ Rev.\ D {\bf 90}, no. 1, 015029 (2014)
  [arXiv:1404.6599 [hep-ph]];
   K.~K.~Boddy and J.~Kumar,
  Phys.\ Rev.\ D {\bf 92}, no. 2, 023533 (2015)
  [arXiv:1504.04024 [astro-ph.CO]].

\bibitem{Gonzalez-Morales:2014eaa} 
  A.~X.~Gonzalez-Morales, S.~Profumo and F.~S.~Queiroz,
  Phys.\ Rev.\ D {\bf 90}, no. 10, 103508 (2014)
  [arXiv:1406.2424 [astro-ph.HE]].
  
\bibitem{Ackermann:2013yva} 
  M.~Ackermann {\it et al.} [{\it Fermi}-LAT Collaboration],
  Phys.\ Rev.\ D {\bf 89}, 042001 (2014)
  [arXiv:1310.0828 [astro-ph.HE]].

\bibitem{Bonnivard:2015tta} 
  V.~Bonnivard {\it et al.},
  Astrophys.\ J.\  {\bf 808}, no. 2, L36 (2015)
  [arXiv:1504.03309 [astro-ph.HE]].

\bibitem{Ackermann:2015abc} 
M.~Ackermann {\it et al.} [{\it Fermi}-LAT Collaboration],
  arXiv:1503.02641 [astro-ph.HE];


\bibitem{GeringerSameth:2011iw} 
  A.~Geringer-Sameth and S.~M.~Koushiappas,
  Phys.\ Rev.\ Lett.\  {\bf 107}, 241303 (2011)
  [arXiv:1108.2914 [astro-ph.CO]].
  A.~Geringer-Sameth, S.~M.~Koushiappas and M.~G.~Walker,
  Phys.\ Rev.\ D {\bf 91}, no. 8, 083535 (2015)
  [arXiv:1410.2242 [astro-ph.CO]].
  
 
\bibitem{Walker:2009zp} 
  M.~G.~Walker, M.~Mateo, E.~W.~Olszewski, J.~Penarrubia, N.~W.~Evans and G.~Gilmore,
  Astrophys.\ J.\  {\bf 704}, 1274 (2009)
  [Astrophys.\ J.\  {\bf 710}, 886 (2010)]
  [arXiv:0906.0341 [astro-ph.CO]].

\bibitem{Wolf:2009tu} 
  J.~Wolf, G.~D.~Martinez, J.~S.~Bullock, M.~Kaplinghat, M.~Geha, R.~R.~Munoz, J.~D.~Simon and F.~F.~Avedo,
  Mon.\ Not.\ Roy.\ Astron.\ Soc.\  {\bf 406}, 1220 (2010)
  [arXiv:0908.2995 [astro-ph.CO]].
  
\bibitem{Martinez:2013els} 
  G.~D.~Martinez,
  arXiv:1309.2641 [astro-ph.GA].

\bibitem{Bechtol:2015cbp} 
  K.~Bechtol {\it et al.} [DES Collaboration],
  Astrophys.\ J.\  {\bf 807}, no. 1, 50 (2015)
  [arXiv:1503.02584 [astro-ph.GA]].
  
\bibitem{Koposov:2015cua} 
  S.~E.~Koposov, V.~Belokurov, G.~Torrealba and N.~W.~Evans,
  Astrophys.\ J.\  {\bf 805}, no. 2, 130 (2015)
  [arXiv:1503.02079 [astro-ph.GA]]. 
  
\bibitem{Laevens:2015una} 
  B.~P.~M.~Laevens {\it et al.},
  Astrophys.\ J.\  {\bf 802}, L18 (2015)
  [arXiv:1503.05554 [astro-ph.GA]].

\bibitem{DESCollaboration}
DES Collaboration, [arXiv:1508.03622 [astro-ph.GA]].
     
  
\bibitem{Geringer-Sameth:2014yza} 
  A.~Geringer-Sameth, S.~M.~Koushiappas and M.~Walker,
  Astrophys.\ J.\  {\bf 801}, no. 2, 74 (2015)
  [arXiv:1408.0002 [astro-ph.CO]].
    
\bibitem{Fermiback} \url{http://fermi.gsfc.nasa.gov/ssc/data/access/lat/BackgroundModels.html}  

  \bibitem{Cirelli:2010xx} 
  M.~Cirelli {\it et al.},
  JCAP {\bf 1103}, 051 (2011)
  [JCAP {\bf 1210}, E01 (2012)]
  [arXiv:1012.4515 [hep-ph], arXiv:1012.4515 [hep-ph]].
  
  \bibitem{Sjostrand:2007gs} 
  T.~Sjostrand, S.~Mrenna and P.~Z.~Skands,
  Comput.\ Phys.\ Commun.\  {\bf 178}, 852 (2008)
  [arXiv:0710.3820 [hep-ph]].  

\bibitem{Dwarfparameters}
Gregory D. Martinez,
Monthly Notices of the Royal Astronomical Society 2015 451 (3): 2524-2535,  arXiv:1309.2641.


\bibitem{Strigari:2007at} 
  L.~E.~Strigari, S.~M.~Koushiappas, J.~S.~Bullock, M.~Kaplinghat, J.~D.~Simon, M.~Geha and B.~Willman,
  Astrophys.\ J.\  {\bf 678}, 614 (2008)
  [arXiv:0709.1510 [astro-ph]].
  
\bibitem{Rolke:2004mj} 
  W.~A.~Rolke, A.~M.~Lopez and J.~Conrad,
  Nucl.\ Instrum.\ Meth.\ A {\bf 551}, 493 (2005)
  doi:10.1016/j.nima.2005.05.068
  [physics/0403059].
  

\bibitem{Bolz:2000fu} 
  M.~Bolz, A.~Brandenburg and W.~Buchmuller,
  Nucl.\ Phys.\ B {\bf 606}, 518 (2001)
  [Nucl.\ Phys.\ B {\bf 790}, 336 (2008)]
  [hep-ph/0012052].
  
\bibitem{Pradler:2006qh} 
  J.~Pradler and F.~D.~Steffen,
  Phys.\ Rev.\ D {\bf 75}, 023509 (2007)
  [hep-ph/0608344].  
  
  
  

\bibitem{Giesen:2015ufa} 
  G.~Giesen, M.~Boudaud, Y.~Genolini, V.~Poulin, M.~Cirelli, P.~Salati and P.~D.~Serpico,
  arXiv:1504.04276 [astro-ph.HE].  
          
 \bibitem{Audren:2014bca} 
  B.~Audren, J.~Lesgourgues, G.~Mangano, P.~D.~Serpico and T.~Tram,
  JCAP {\bf 1412}, no. 12, 028 (2014)
  [arXiv:1407.2418 [astro-ph.CO]].
  
 \bibitem{Aisati:2015vma} 
  C.~E.~Aisati, M.~Gustafsson and T.~Hambye,
  arXiv:1506.02657 [hep-ph];
  L.~Covi, M.~Grefe, A.~Ibarra and D.~Tran,
  JCAP {\bf 1004}, 017 (2010)
  [arXiv:0912.3521 [hep-ph]];
  S.~Palomares-Ruiz,
  Phys.\ Lett.\ B {\bf 665}, 50 (2008)
  [arXiv:0712.1937 [astro-ph]];
  A.~Esmaili, A.~Ibarra and O.~L.~G.~Peres,
  JCAP {\bf 1211}, 034 (2012)
  [arXiv:1205.5281 [hep-ph]].  

\bibitem{Geringer-Sameth:2015lua} 
  A.~Geringer-Sameth, M.~G.~Walker, S.~M.~Koushiappas, S.~E.~Koposov, V.~Belokurov, G.~Torrealba and N.~W.~Evans,
  Phys.\ Rev.\ Lett.\  {\bf 115}, no. 8, 081101 (2015)
  [arXiv:1503.02320 [astro-ph.HE]].
  
\bibitem{Drlica-Wagner:2015xua} 
  A.~Drlica-Wagner {\it et al.} [Fermi-LAT and DES Collaborations],
  Astrophys.\ J.\  {\bf 809}, no. 1, L4 (2015)
  [arXiv:1503.02632 [astro-ph.HE]].  

\bibitem{Hooper:2015ula} 
  D.~Hooper and T.~Linden,
  JCAP {\bf 1509}, no. 09, 016 (2015)
  [arXiv:1503.06209 [astro-ph.HE]].
  




  
\end{thebibliography}
\end{document}